# XPS STUDY OF THE CHEMICAL STABILITY OF $DyBa_2Cu_3O_{6+\delta}$ SUPERCONDUCTOR

A.V. Fetisov[1], G.A. Kozhina[1], S.Kh. Estemirova[1], V.B. Fetisov[2], R.I. Gulyaeva[1]

[1]*Institute of Metallurgy, Ural Branch of the Russian Academy of Sciences,101 Amundsen str., 620016 Ekaterinburg, Russia*
[2] *Ural State Agrarian University, 42Karl Liebknecht str., Ekaterinburg, 620075, Russia*

Correspondence should be addressed to Galina Kozhina, gakozhina@yandex.ru

**Abstract**

The chemical stability of the powder $DyBa_2Cu_3O_{6+\delta}$ has been studied by X-ray diffraction (XRD), X-ray photoelectron spectroscopy (XPS) and thermal analysis at ambient conditions. The powder was subjected to mechanical processing in a ball mill-activator to accelerate chemical degradation. The kinetic regularities of hydrolytic decomposition of $DyBa_2Cu_3O_{6+\delta}$ under the influence of air moisture have been determined. The resistive properties of $DyBa_2Cu_3O_{6+\delta}$ to water have been found to be better, but not much different from analogous properties of $YBa_2Cu_3O_{6+\delta}$ which is unstable in a wet environment. Chemical degradation of the material is triggered by crucial concentrating of water particles near the free surface of the solid reactant (due to their low diffusibility in the bulk) leading to rapid chemical decomposition of the respective regions.

## 1. Introduction

Compounds $REBa_2Cu_3O_{6+\delta}$ (RE = rare earth element) are well known as high-temperature superconductors (HTSC) with the temperature of the transition to the superconducting state of 90–93 K [1]. At liquid nitrogen temperature and in high fields the intrinsic current-carrying properties of $REBa_2Cu_3O_{6+\delta}$ are superior to those of Bi, Tl and Hg superconducting phases [1, 2]. The composition with RE = Y is a conventional material for superconducting electronics under development [3, 4], magnetic field sensors of $\pi$ SQUID type [3], high current cables [5], and other products. On the other side, low chemical stability of most members of these HTSC-oxides restricts its applicability as functional materials [1, 3]. This fact is the main cause of great number of studies on the interaction of $REBa_2Cu_3O_{6+\delta}$ with water and moist air.

As a result, it was found that surface of these oxides has polar properties; this entails a very high affinity of it for the polar molecules, such as $H_2O$ [6]. Nonetheless, moisture-resistive properties of these HTSC depend significantly on the type of RE. So, high chemical stability of $NdBa_2Cu_3O_{6+\delta}$ and relatively small chemical stability of $YBa_2Cu_3O_{6+\delta}$ was noted [7–10]. In an attempt to explain the differences in properties of the HTSC oxides, authors of [7–10] pay attention to the cationic nonstoichiometry inherent to $NdBa_2Cu_3O_{6+\delta}$: the synthesis usually results in a solid solution $Nd_{1+x}Ba_{2-x}Cu_3O_{6+\delta}$ with excess neodymium ions in structural positions of barium (if the precursors are taken based on the stoichiometric composition, then impurity phases $Ba_2Cu_3O_{5+y}$ and $BaCO_3$ are formed together with the solid solution). Hypothetical situations related to the cation redistribution (disordering of the oxygen sublattice [9], the strengthening of bonds between oxygen and the lattice [10] and others) are assumed to lead to moisture-resistive properties.

In [11], we compared the chemical stability of $YBa_2Cu_3O_{6+\delta}$ and $NdBa_2Cu_3O_{6+\delta}$ under real atmospheric conditions. The mechanical activation method was used to accelerate the environmental degradation proceeding during investigation. We have obtained the result, which do not contradict the main experimental finding of works [7–10]: $NdBa_2Cu_3O_{6+\delta}$ exhibits enhanced corrosion resistance against atmospheric moisture compared with $YBa_2Cu_3O_{6+\delta}$. It has been suggested that enhancing of the resistive properties of $NdBa_2Cu_3O_{6+\delta}$ is largely promoted by flowing its hydrolytic decomposition in a special way, non-traditional for $REBa_2Cu_3O_{6+\delta}$. The main idea is that hydrolysis in this case occurs without formation of so-called "color" phase with the general formula $RE_2BaCuO_5$ that is non effective in terms of energy barriers. In the case of $YBa_2Cu_3O_{6+\delta}$, the ability to form this phase makes the process of decomposition easier.

In this work, we have tried to understand the importance of the aforementioned factor of chemical stability and its applicability to one more member of the oxide family $REBa_2Cu_3O_{6+\delta}$: $DyBa_2Cu_3O_{6+\delta}$. The motivation of this study is data [12] on high chemical stability of the $DyBa_2Cu_3O_{6+\delta}$ compound by comparison with $YBa_2Cu_3O_{6+\delta}$ (that makes it similar to $NdBa_2Cu_3O_{6+\delta}$), while its superconducting performance characteristics ($T_c$, $J_c$, dependence of critical current on external magnetic fields) are the same as for $YBa_2Cu_3O_{6+\delta}$ [2]. Our experiments have revealed that "color" phase $Dy_2BaCuO_5$ is completely absent on the surface of mechanically activated $DyBa_2Cu_3O_{6+\delta}$ and decomposition products are almost the same as in the case of neodymium-barium cuprate. Nevertheless, kinetics of the hydrolysis process for $DyBa_2Cu_3O_{6+\delta}$ and $NdBa_2Cu_3O_{6+\delta}$ is significantly different each from other, which has led to a revision of our initial assumptions.

## 2. Experimental technique

The compound $DyBa_2Cu_3O_{6+\delta}$ was obtained by stepwise annealing a mixture of barium carbonate $BaCO_3$ ("chemically pure" grade) and oxides of copper and dysprosium ("especially pure" grade) at 950°C (100 h). The product was subjected to oxidative annealing by slow (with furnace) cooling from the synthesis temperature to 400°C and exposed to this temperature in air for 3 hours. The oxygen content of the sample ($6+\delta$) according to the oxidative annealing conditions can be estimated as 6.90±0.03. The synthesis of the $Dy_2BaCuO_5$ was performed at 1000°C for 100 hours, using $Dy_2O_3$, $BaCO_3$ and pre-synthesized $DyBa_2Cu_3O_{6+\delta}$ as precursors.

Mechanical activation of the synthesized $DyBa_2Cu_3O_{6+\delta}$ was performed in an AGO-2 planetary ball mill (set material – Fe) with acceleration of milling bodies of 60 g. The as-synthesized powders were dry milled for 2 and 10 min. The particles size was determined using a scanning electron microscope Carl Zeiss EVO 40 (Germany) and amounted to $(3 \div 5) \cdot 10^{-6}$ m. While the particles themselves were agglomerates consisting of stuck together crystallites with the size of nearly $0.3 \cdot 10^{-6}$ m.

X-ray examination of samples was carried out on a Shimadzu XRD-7000 diffractometer in CuKα-radiation using certified silicon powder as an internal standard. Measurement conditions: I = 30 mA, V = 40 kV, the angle range from 20° to 70° in 2Θ, a step size of 0.03° and a step scan of 2 s.

Thermal studies of $DyBa_2Cu_3O_{6+\delta}$ were carried out using the thermal analyser Netzsct STA 449C Jupiter (Germany), combined with a mass spectrometer Netzsct QMS 403C Aёolos (Germany) for evolved gas analysis. The rates of heating and cooling of samples were 10 K/min.

XPS study was performed on an Omicron Multiprob spectrometer (Germany) equipped with an electron analyzer EA 125. The source of the exciting radiation for XPS was Mg-anode of X-ray gun with 170 W. The energy scale of the spectrometer was calibrated by the binding energies of the peaks: Au $4f_{7/2}$ (84.00 eV), Ag $3d_{5/2}$ (368.29 eV) and Cu $2p_{3/2}$ (932.67 eV). The value of samples surface charge was determined by the difference between the measured binding energy of the C 1$s$ signal due to C-H groups and the value of 285.0 eV. High-resolution spectrum was recorded with 0.05 eV energy step, exposure time of 7.5 s (15 scans by 0.5 s) and an analyser pass energy of 20 eV. The accuracy of peak position determination was 0.1 eV. Decomposition of the spectra into components and extraction of the background were performed by the Shirley method using XPSPeak 4.1 Software. The analysis was carried out on the basis of mutual quantitative consistency between the spectra of different elements belonging to the one compound in accordance with their stoichiometric ratios.

## 3. Experimental results

### 3.1. $Dy_2BaCuO_5$

X-ray diffraction analysis revealed that the sample is single-phase and has the orthorhombic Pnma structure with following lattice parameters: $a$ = 12.2189(7) Å, $b$ = 5.6793(4) Å and $c$ = 7.1523(5) Å. The results of XPS studies of $Dy_2BaCuO_5$ are shown in Fig. 1 and Table 1. It should be noted that in the course of decomposition of the spectra, along with the lines belonging to different phases at the surface some signals corresponding to unrealistic low binding energies (shown by hatching in Fig. 1) were found (for example, we have found the lowest binding energy of O 1s electron level to be near 526 eV, whereas the lowest binding energy of O 1s for metal oxides is considered to be not lower than 528 eV [13, 14]). Consequently, these signals were interpreted as "double charge" effect.

As shown in fig. 1a, the spectrum of oxygen, O 1s, in addition to the satellites resulting from the charging effect, contains two real peaks at 529.3 and 531.5 eV. In similar cases, is considered that the low-energy component of O 1s spectrum corresponds to basic oxide, and the high-energy one is attributed to contaminations, which always present on the surface of samples [14]. Indeed, the binding energy of 531.5 eV is in good agreement with the data on the carbonate group, $CO_3^{2-}$, located on the surface of various oxides [14].

The spectra of barium and copper, Ba 3$d$ and Cu $2p_{3/2}$, represent two single peaks at 779.4 eV and 932.9, respectively (see figures 1b and 1c). We believe it to be attributed to the main phase, having, however, in mind the possibility of a small carbonate component in the Ba 3$d$ spectrum (with the Ba $3d_{5/2}$ peak being centered at ~780.0 eV [11]), which is not noticeable due to its insignificance. In turn, in the Dy $3d_{5/2}$ spectrum (see fig. 1d) we can see a low-energy peak at 1294.5 eV relating, as we believe, to the main phase and a high-energy peak at 1296.4 eV corresponding to $Dy_2O_3$ [15].

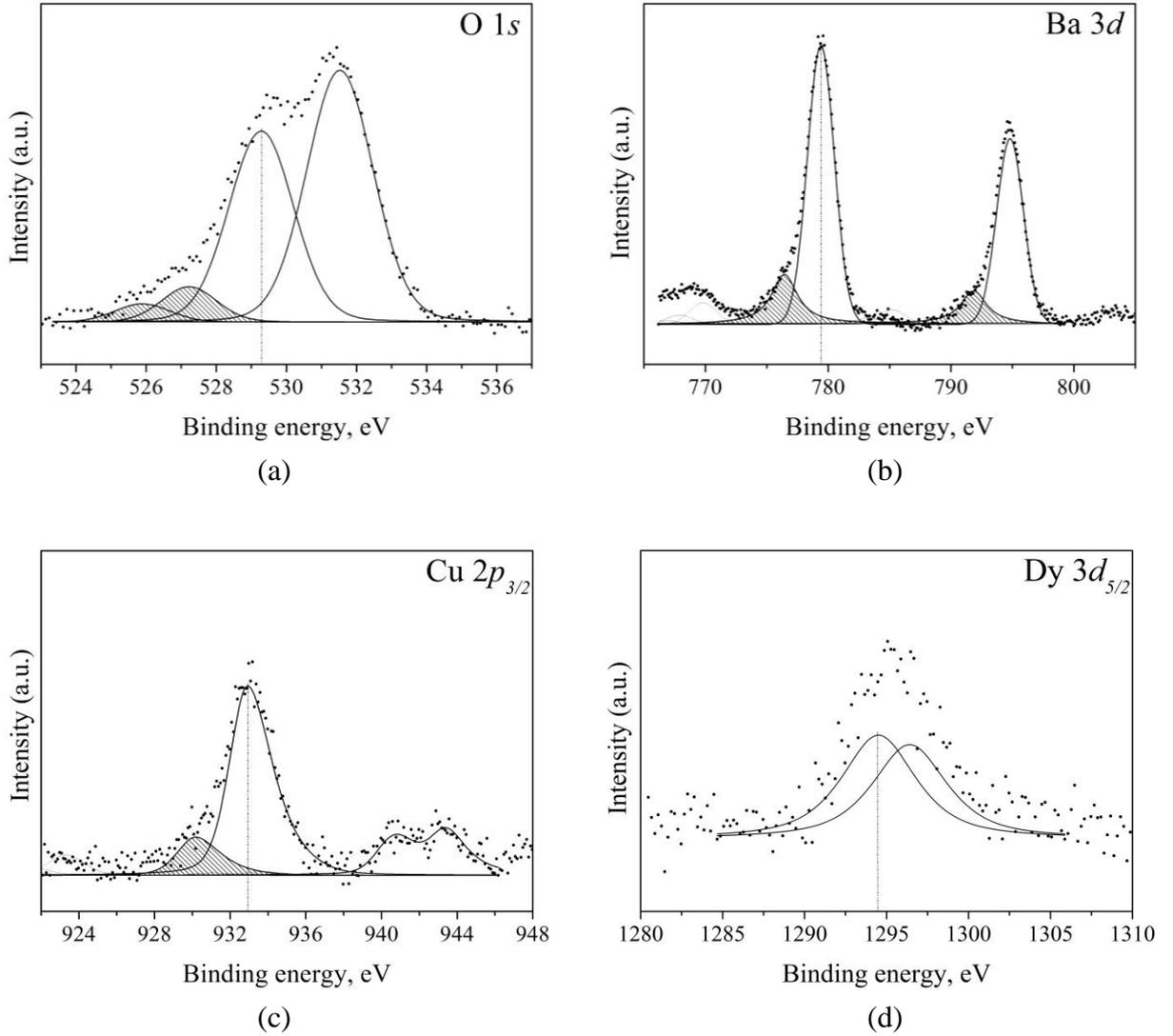

**Fig. 1.** O 1$s$, Ba 3$d$, Cu 2$p_{3/2}$ and Dy 3$d_{5/2}$ XPS spectra of Dy$_2$BaCuO$_5$.

**Table 1:** XPS data for DyBa$_2$Cu$_3$O$_{6+\delta}$.

| Electron level | Binding energy, eV |
|---|---|
| O 1$s$ | 529.3 |
| Ba 3$d$ | 779.4 |
| Cu 2$p$ | 932.9 |
| Dy 3$d$ | 1294.5 |

### 3.2. DyBa$_2$Cu$_3$O$_{6+\delta}$

X-ray diffraction study of the as-synthesized and two mechanically activated powders of DyBa$_2$Cu$_3$O$_{6+\delta}$ shows that all these powders have the orthorhombic Pmmm structure with little difference in lattice parameters (see Table 2). However, the mechanically activated samples, in contrast to the as-synthesized micropowder, contain some impurities: there are additional low-intensity peaks on their diffractograms in the angles range 2θ = 28.5–30.5 degrees. Due to the significant width of the peaks and the presence of high background noise, unambiguous interpretation of impurity phases on the obtained diffractograms is difficult. Both the "color" phase $Dy_2BaCuO_5$ and the barium cuprate $Ba_2Cu_3O_{5+y}$ can be the impurity (see Fig. 2, the X-ray line-diagramms of these phases are taken from the ICDD database).

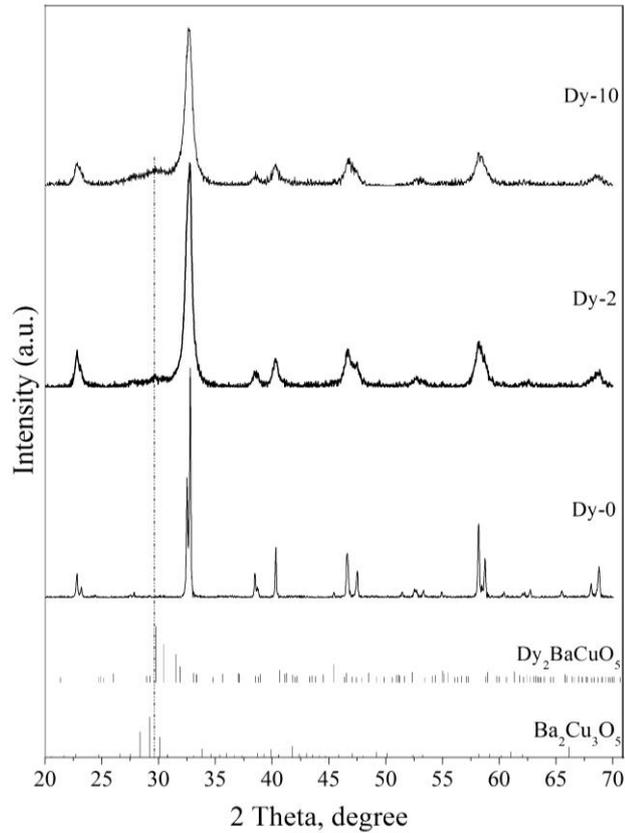

**Fig. 2.** X-ray diffraction patterns of $DyBa_2Cu_3O_{6+\delta}$ samples subjected to mechanical treatment of different duration.

**Table 2:** Phase composition and lattice parameters of $DyBa_2Cu_3O_{6+\delta}$ before and after mechanical treatment

| Crystallographic parameters | Milling duration in AGO-2 mill | | |
|---|---|---|---|
| | 0-min (sample Dy-0) | 2 min (sample Dy-2) | 10min (sample Dy-10) |
| The main phase - orthorhombic S.G. Pmmm | | | |
| $a$, Å | 3.8270(3) | 3.8297(7) | 3.8336(12) |
| $b$, Å | 3.8928(7) | 3.8934(12) | 3.8940(23) |
| $c$, Å | 11.6926(16) | 11.6806(39) | 11.6835(78) |
| $V$, Å$^3$ | 174.197(64) | 174.162(145) | 174.412(274) |
| Impurity, mass% | – | ~3% | ~ 5% |

Refinement of the phase composition of samples was carried out by XPS. Corresponding XPS-spectra are shown in Figure 3. Binding energies of the characteristic electron levels obtained by analyzing the spectra are listed in Table 3.

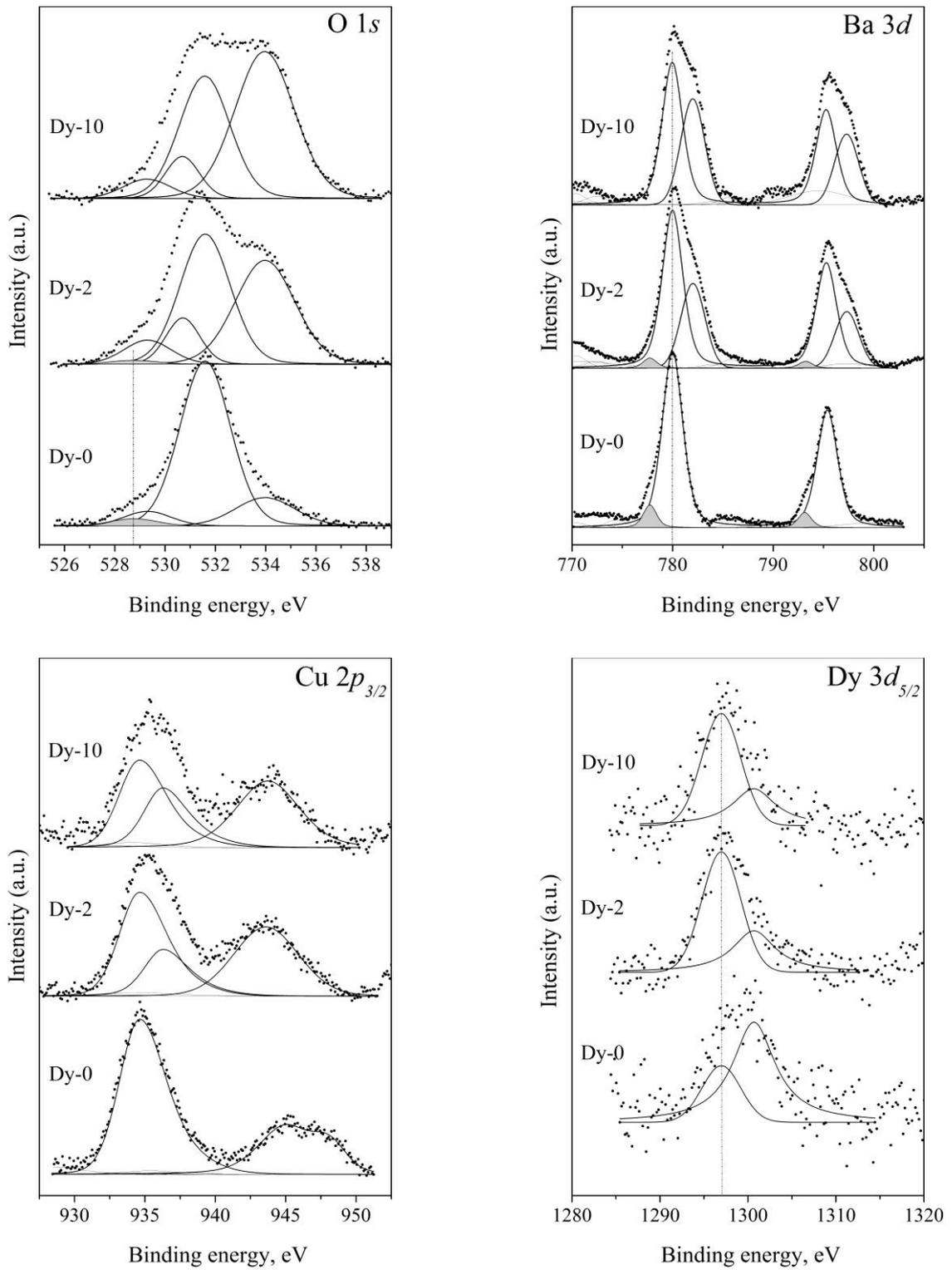

**Fig. 3.** O 1$s$, Ba 3$d$, Cu 2$p_{3/2}$ and Dy 3$d_{5/2}$ XPS spectra of the as-synthesized and mechanoactivated powders of DyBa$_2$Cu$_3$O$_{6+\delta}$. As the graph labelling here the naming of the appropriate samples in Table. 2 has been used.

**Table 3**: XPS data for DyBa$_2$Cu$_3$O$_{6+\delta}$.

| Electron level | Binding energy, eV | | |
|---|---|---|---|
| | Dy-0 | Dy-2 | Dy-10 |
| O1$s$ | **528.8*** | **528.8** | - |

|  |  |  |  |  |
|---|---|---|---|---|
|  |  | 529.3 | 529.3 | 529.3 |
|  |  | - | 530.7 | 530.7 |
|  |  | 531.6 | 531.6 | 531.6 |
|  |  | 534.0 | 534.0 | 534.0 |
| Ba $3d_{5/2}$ |  | **777.7** | **777.7** | - |
|  |  | 780.0 | 780.0 | 780.0 |
|  |  | - | 782.0 | 782.0 |
| Cu $2p_{3/2}$ |  | 934.6 | 934.6 | 934.6 |
|  |  | - | 936.3 | 936.3 |
| Dy $3d_{5/2}$ |  | 1296.7 | 1296.7 | 1296.7 |
|  |  | 1300.7 | 1300.7 | 1300.7 |

*peaks directly related to DyBa$_2$Cu$_3$O$_{6+\delta}$ are in bold

When analyzing the presented O 1$s$, Ba 3$d$, and Dy 3$d$ spectra, it is not difficult to identify clearly distinguishable peaks at 531.6, 780.0, and 1296.7 eV, respectively. So, according to [11], the two forward peaks are attributed to the well-studied compound BaCO$_3$, and, in turn, the latter one is related [15] with the presence of Dy$_2$O$_3$ on the surface of DyBa$_2$Cu$_3$O$_{6+\delta}$. Besides, during deconvolution of the O 1$s$ spectrum, it should be also taken into account that dysprosium oxide has the oxygen peak at 529.3 [15]. At the same time in the O 1$s$ and Ba 3$d$ spectra one can see slight peaks at 528.8 and 777.7 eV which we interpret as a contribution of the main phase (shown by tinting) due to the proximity of the binding energy of corresponding peaks in the compounds YBa$_2$Cu$_3$O$_{6+\delta}$ and NdBa$_2$Cu$_3$O$_{6+\delta}$ [11, 16–19]. To allocate the DyBa$_2$Cu$_3$O$_{6+\delta}$ contributions in the Cu 2$p$ and Dy 3$d$ spectra is not possible.

The group of three lines in O 1$s$, Cu 2$p_{3/2}$, and Ba 3$d_{3/2}$ spectra at 530.7, 936.2, and 781.9 eV, respectively, whose intensities are synchronously changing in the course of the mechanical processing time (the last parameter is marked in Fig. 3 as number N in Dy-N, that means duration in minutes), obviously belongs to the same compound. Exactly the same set of synchronously changing lines was obtained in our previous study of mechanically activated NdBa$_2$Cu$_3$O$_{6+\delta}$ [11]; using X-ray diffraction analysis it was concluded that these lines correspond to double oxide Ba$_2$Cu$_3$O$_{5+y}$.

In the work [11], the presence of some quantities of the RE oxalates in YBa$_2$Cu$_3$O$_{6+\delta}$ and NdBa$_2$Cu$_3$O$_{6+\delta}$ was revealed after mechanical activation of the samples. Likewise, the peak at 1300.7 eV in the Dy 3$d$ spectra perhaps is related to Dy$_2$(C$_2$O$_4$)$_3$; according to quantitative analysis, the oxygen peak of this compound likely to be located around 531.5 eV and superimposed on the peak of carbonate groups.

Further, the oxygen peak at 534.0 eV can not be attributed to the structure of any above mentioned compounds on account of the extra-large binding energy value. Presumably, it is a consequence of organic contaminants adsorbed on the surface. Such conclusion is typical for XPS studies (see, e.g., [14, 20]).

Lastly, it should be particularly noted that the binding energy of the Cu 2$p_{3/2}$ main peak (934.6 eV) is not characteristic for this family of HTSC cuprates. In most cases, in XPS studies devoted to REBa$_2$Cu$_3$O$_{6+\delta}$, the binding energy of Cu2$p_{3/2}$ is much smaller (by ~1 eV). This value corresponds either directly to the HTSC phase (933.5 eV [18, 21]) or to the product of its hydrolysis – CuO (933.7±0.1 эВ [22–24]). For instance, in our work [11], the Cu 2$p_{3/2}$ main peak of YBa$_2$Cu$_3$O$_{6+\delta}$ (Y-0, Y-2, Y-10) and NdBa$_2$Cu$_3$O$_{6+\delta}$ (Nd-0, Nd-2, Nd-10) specimens was revealed at $E_b$ = 933.6±0.1 eV. The bias of the Cu2$p_{3/2}$ peak of DyBa$_2$Cu$_3$O$_{6+\delta}$ to high energies is especially notable in direct comparison shown in Fig. 4. Based on literature data, we came to the conclusion that hydroxide, carbonate and basic carbonate of copper could equally be responsible for the increased value of the Cu 2$p_{3/2}$-binding energy of dysprosium HTS samples (corresponding values of binding energy are 934.75±0.05 eV [22, 24–26]).

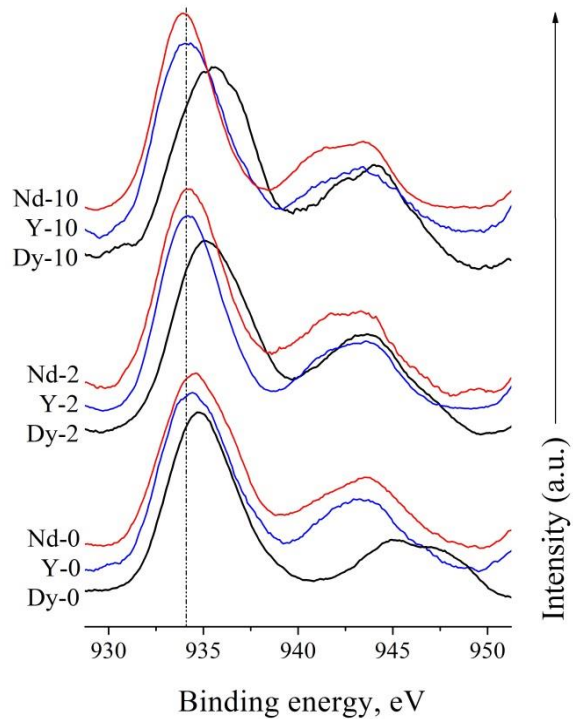

Fig. 4. Comparison of the Cu 2$p_{3/2}$ peaks of REBa$_2$Cu$_3$O$_{6+\delta}$; RE = Dy, Y, Nd (original data have been subjected to a smoothing procedure).

Thus, the XPS data supplement the XRD data: barium cuprate Ba$_2$Cu$_3$O$_{5+y}$ is the impurity phase in mechanically activated DyBa$_2$Cu$_3$O$_{6+\delta}$ powder; no trace of the "color" phase in the XPS spectra of these powders has been found. Moreover, the DyBa$_2$Cu$_3$O$_{6+\delta}$ surface turns out to contain some copper compound as a chemical degradation product (with the Cu 2$p_{3/2}$ binding energy of 934.6 eV) which is not characteristic for REBa$_2$Cu$_3$O$_{6+\delta}$. Further, by means of thermal analysis we'll make an attempt to specify the chemistry of this compound.

The features of gas exchange occurring between DyBa$_2$Cu$_3$O$_{6+\delta}$ powders and the gas phase detected by thermal analysis at elevated temperatures are shown in Fig. 5. For comparison, similar data obtained on mechanically activated NdBa$_2$Cu$_3$O$_{6+\delta}$ powders (taken from our paper [11]) are presented on the same figure.

The thermogravimetric curve obtained for the as-synthesized DyBa$_2$Cu$_3$O$_{6+\delta}$ sample shows the behavior typical for the entire family of REBa$_2$Cu$_3$O$_{6+\delta}$ superconductors (see Fig. 5a). In particular, this is clearly seen when comparing it with the TG curve obtained for the similar NdBa$_2$Cu$_3$O$_{6+\delta}$ sample. It is well known that weight changes of REBa$_2$Cu$_3$O$_{6+\delta}$ under appropriate heating conditions is associated with oxygen desorption [1, 27]. In the mechanically activated samples, however, the additional weight loss in ranges of 350–550 and 800–900°C is observed.

According to mass spectrometry (Fig. 5b), it is due to escape of moisture and carbon dioxide, respectively.

As known, mass-spectrometry has higher sensitivity than TGA and, consequently, allows getting a more detailed analysis of the thermal decomposition of $DyBa_2Cu_3O_{6+\delta}$. According to it, the main water vapor flow is actually escaping from $DyBa_2Cu_3O_{6+\delta}$ in a range of 350–650°C, where one can see a clear Gaussian peak with a maximum at ~495°C in Fig. 5b. Given the narrow range and the high value of temperature revealed for thermal desorption, one can be assumed that during mechanical activation of $DyBa_2Cu_3O_{6+\delta}$ its structure acquires coordinated water. With that, the mass-spectrum contains, in addition, a small $H_2O$-peak at 105–245°C, which is accompanied by the barely noticeable decrease in the sample weight (see Fig. 5a). In the section devoted to XPS analysis (see above), we found the presence of one or more copper compounds containing $OH^-$ and/or $CO_3$ groups (they haven't been identified by spectrometry) on the surface of $DyBa_2Cu_3O_{6+\delta}$. Since the low-temperature water peak of the mass-spectrum is not accompanied by any $CO_2$-signal (see Fig. 5b), then copper hydroxide, $Cu(OH)_2$, is the most likely source of it. To confirm this hypothesis, we conducted thermal analysis of freshly-synthesized copper hydroxide (see the callout in Fig. 5b), which showed the decomposition temperature of this compound under our experimental conditions[*] to be 134°C[**]. It is quite close to the dehydration temperature of Dy-10 (150°C). The latter is a thin film disposed on HTSC surface and it is exposed by the adhesive force from the "substrate". May be, that is why, it has a little elevated temperature of the dehydration.

The thermal analysis presented herein thus serves as a refinement of the phase composition of the $DyBa_2Cu_3O_{6+\delta}$ surface, made above by XPS. Also the regularities of gas exchange for dysprosium HTS was found to differ radically from those observed for $NdBa_2Cu_3O_{6+\delta}$. Desorption of water from the latter proceeds mainly in a temperature range of 50–370°C (see Fig. 5b) which is typical of dissolved water.

The intensive signal of $CO_2$ observed in a high temperature (above ~770°C) corresponds to the decomposition reaction of $BaCO_3$ under our experimental conditions (according to passport data, $CO_2$ content in the air used for blowing of thermoanalyzer is not more than $2 \cdot 10^{-4}$ vol%).

---

[*] The heating rate and atmosphere were the same as they were in the study of sample Dy-10
[**] Amorphous $Cu(OH)_2$ was obtained by mixing the $CuCl_2$ and NaOH solutions, washing precipitated cupric hydroxide, and drying it by heating to 50°C. A value of the amorphous hydroxide decomposition temperature found to be 134 °C that corresponds to the expected one (it is well known that decomposition of this compound starts from 70–90°C). At the same time, according to XRD analysis the resulting product also contained $Cu(OH)_2$ as a crystalline phase. It explains the presence of a narrow peak at 197°C in the thermogram of the compound.

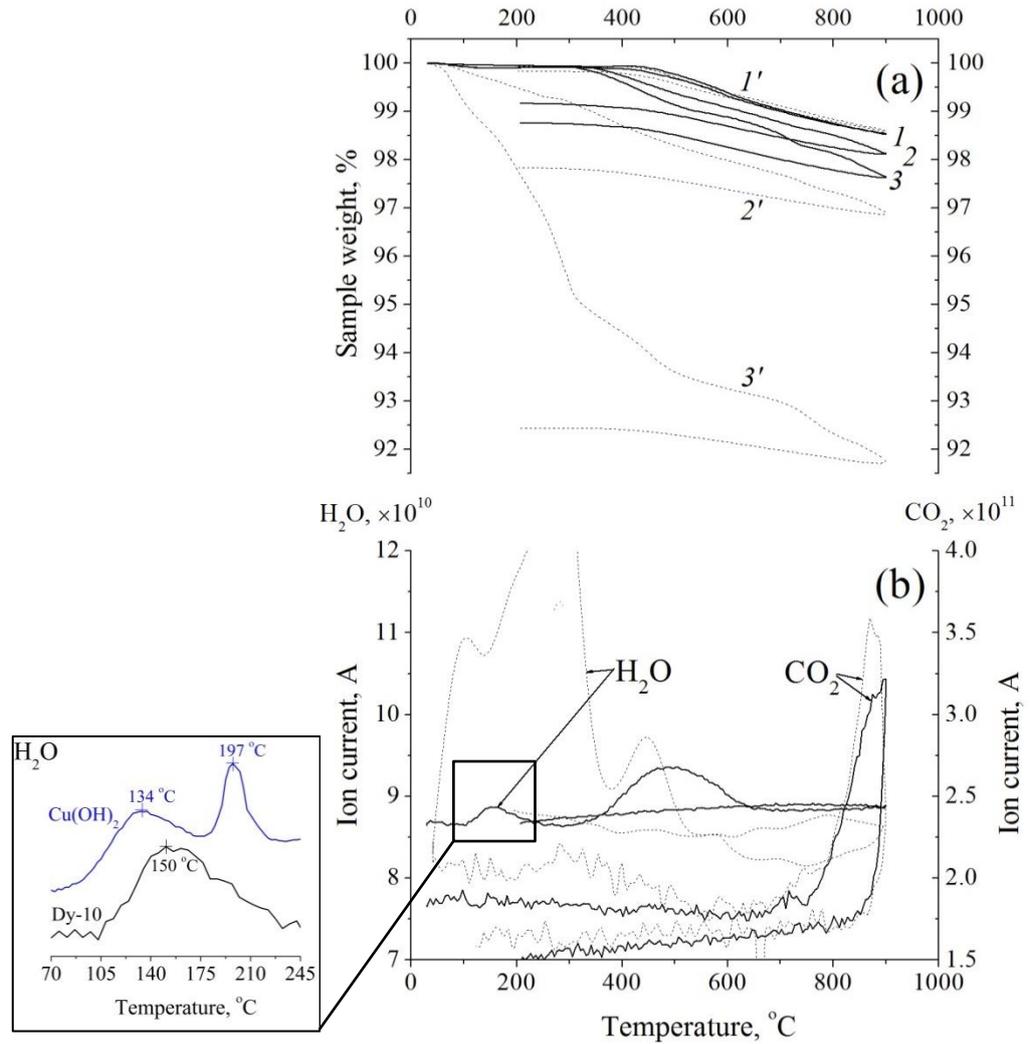

**Fig. 5.** Thermal analysis results.

(a) Thermogravimetric curves for the samples: Dy-0 (1), Nd-0 (1'); Dy-2 (2), Nd-2 (2'), Dy-10 (3) and Nd-10 (3').
(b) Mass spectrometry data for the samples: Dy-10 (solid curves) and Nd-10 (dashed curves). $H_2O$ evolution curve for the dysprosium sample is biased up by $7.5·10^{-10}$ A for convenience.
In callout: The element of the mass spectrometric curve of the Dy-10 sample ($H_2O$-signal) is shown together with the corresponding curve of $Cu(OH)_2$.

## 4. Discussion of Results

In the experimental part of the work, using the XPS-analysis we have proved that no traces of the "color" phase $Dy_2BaCuO_5$ is on the surface of mechanically activated $DyBa_2Cu_3O_{6+\delta}$ samples; but instead of this the following compounds are formed as products of hydrolysis: $Ba_2Cu_3O_{5+\beta}$, $Ba(OH)_2$ (as barium hydroxide reacts readily with ambient carbon dioxide, the product $BaCO_3$ was recorded instead), $Cu(OH)_2$, and $Dy_2O_3$ (partially transformed to $Dy_2(C_2O_4)_3$). So, the mechanism of chemical decomposition of the oxide, at least under mechanical activation, differs significantly from that of YBa$_2$Cu$_3$O$_{6+\delta}$[*]. With that, considered in [11] hydrolysis of NdBa$_2$Cu$_3$O$_{6+\delta}$, occurring during mechanical activation, led to approximately the same phase composition of the reaction products as in the case of DyBa$_2$Cu$_3$O$_{6+\delta}$ examined here. The only difference is the presence of CuO (instead of cupric hydroxide) in neodymium cuprate samples. Hence, the chemistry of decomposition of both oxides under the action of water is similar.

Despite the differences in chemistry, the influence of moisture on DyBa$_2$Cu$_3$O$_{6+\delta}$ is almost as destructive as on YBa$_2$Cu$_3$O$_{6+\delta}$: the latter contained impurity phases in amount of 7% after 10 minutes mechanical activation [11] versus 5% in the case of dysprosium-barium cuprate. That does not confirm the findings of other authors [12] about high chemical stability of DyBa$_2$Cu$_3$O$_{6+\delta}$. The process of chemical degradation of NdBa$_2$Cu$_3$O$_{6+\delta}$ under the same conditions was much less intense: X-ray diffraction did not record impurity phases, and XPS-analysis showed traces of the main phase even after 10 min mechanical activation.

The inverse power law dependence of the hydrolysis degree of DyBa$_2$Cu$_3$O$_{6+\delta}$ on the mechanical activation duration (obtained on the basis of TG and XRD analysis) indicates that the process of chemical decomposition of the oxide is localized in the surface layer of its particles, and the decomposition rate is limited by diffusion of the starting reagent (H$_2$O) through the layer of hydrolysis products. This is similar to the kinetic peculiarity of hydrolysis of yttrium-barium cuprate and is not consistent with the feature of the hydrolysis kinetics of NdBa$_2$Cu$_3$O$_{6+\delta}$ [11].

All the above evidences clearly show that *chemistry* and *kinetics* of the hydrolytic decomposition of REBa$_2$Cu$_3$O$_{6+\delta}$ do not linked so definitely as suggested in [11]. DyBa$_2$Cu$_3$O$_{6+\delta}$ and YBa$_2$Cu$_3$O$_{6+\delta}$ undergo different ways of chemical degradation under the action of atmospheric moisture, however, they exhibit similar kinetic regularities and parameters of the process. A diametrically opposite situation arises when comparing oxides DyBa$_2$Cu$_3$O$_{6+\delta}$ and NdBa$_2$Cu$_3$O$_{6+\delta}$: the same type of chemistry of their degradation leads to different its implementations in terms of topochemistry and kinetics.

In that case, what actually has the dominant influence on the kinetics of the hydrolysis process of oxides REBa$_2$Cu$_3$O$_{6+\delta}$?

The fact that very strong saturation of neodymium compounds by dissolved water during its mechanical activation had no effect on the intensification of the hydrolysis process [11] seems quite paradoxical at first sight. In turn, the absence of appreciable amount of dissolved water in the mechanically activated DyBa$_2$Cu$_3$O$_{6+\delta}$ (see thermal analysis data above) did not rid the oxide of relatively fast degradation changes. Such reaction of the oxides on the water contained in them seems strange. Let us consider this fact from other positions. When water molecules enter into the lattice of REBa$_2$Cu$_3$O$_{6+\delta}$ chemical interaction occur only if the local concentration of water (its activity) reaches a certain threshold. According to [28] the experimentally obtained value of this concentration in YBa$_2$Cu$_3$O$_{6+\delta}$ is 1.3-1.5 mass%. Below this concentration, water is in the oxide structure in dissolved form (H$_2$O or OH$^-$). So, the actual situation on the development of the hydrolysis is determined by the ratio of the rates of two processes: adsorption of H$_2$O by the oxide surface and the diffusion of water in the crystal lattice. If the diffusion prevails (variant 1) the incoming flow of water particles through the surface is scattered in the oxide bulk. Otherwise (variant 2) water particles are concentrated near the surface, causing chemical decomposition of the material near surface at a certain time. To our opinion, the variant 1 described above could be realized in hydration of the NdBa$_2$Cu$_3$O$_{6+\delta}$ powder and the variant 2 - in the case of DyBa$_2$Cu$_3$O$_{6+\delta}$ (and YBa$_2$Cu$_3$O$_{6+\delta}$). Moreover, in the case of dysprosium-barium cuprate, the variant 2 is implemented with a large overbalance of the rate of adsorption on the diffusion rate. The XPS and thermal analyses data have shown copper hydroxide Cu(OH)$_2$ being the hydrolysis product only of DyBa$_2$Cu$_3$O$_{6+\delta}$, indicating H$_2$O-particle concentration really was increased in the surface layer of this oxide (relative to NdBa$_2$Cu$_3$O$_{6+\delta}$ and YBa$_2$Cu$_3$O$_{6+\delta}$). In the case of NdBa$_2$Cu$_3$O$_{6+\delta}$ and YBa$_2$Cu$_3$O$_{6+\delta}$ [11], water activity in the surface layer was apparently not high enough to transform localized there copper oxide into its hydroxide.

---

[*] Copper hydroxide, barium carbonate and Dy$_2$O$_3$ have not been found in mechanically activated DyBa$_2$Cu$_3$O$_{6+\delta}$-samples by X-ray diffraction as they are apparently in the amorphous state caused by intense mechanical impact. At least, the presence of appreciable amount of Cu(OH)$_2$ and BaCO$_3$ in these samples is confirmed by the mass-spectrometric analysis data (see fig. 4b).

## 5. Conclusion

In this paper, the hydrolysis of the mechanical activated DyBa$_2$Cu$_3$O$_{6+\delta}$ powders has been examined by XPS, XRD and thermal analysis. It has been found that the hydrolysis of DyBa$_2$Cu$_3$O$_{6+\delta}$ results in the formation of Ba(OH)$_2$, Cu(OH)$_2$, Dy$_2$O$_3$, and Ba$_2$Cu$_3$O$_{5+y}$ products on the superconductor surface, with barium hydroxide transforming to carbonate under ambient air conditions. With that, it has not been observed the presence of so-called "color" phase, which is a characteristic product of chemical degradation of same other superconductors, such as YBa$_2$Cu$_3$O$_{6+\delta}$.

During the work the previously suggested idea that the kinetics of hydrolysis of REBa$_2$Cu$_3$O$_{6+\delta}$ is totally determined by the factor of "color phase" (by its presence among the hydrolysis reaction products) was being checked. And this point of view has not been confirmed. The comparative analysis of the experimental results obtained here and in our recent study on NdBa$_2$Cu$_3$O$_{6+\delta}$ and YBa$_2$Cu$_3$O$_{6+\delta}$ samples has shown that decisive role in the chemical degradation kinetics of these oxides is played by the water mobility in the REBa$_2$Cu$_3$O$_{6+\delta}$-structures.

As a general remark it should be noted that although the powders in our study were subjected to mechanical activation treatment, which, as a rule, alters material properties, the hydration process peculiarities obtained are, apparently, real characteristics of oxide under investigation. This conclusion derives from following observations: (a) the XPS peak of copper hydroxide was revealed during the examine of all the samples, including non-activated Dy-0, see figures 3 and 4, whereas copper hydroxide is the indicator of an elevated concentration of moisture in the surface region of DyBa$_2$Cu$_3$O$_{6+\delta}$; (b) the ranking of chemical stability of the HTSC oxides being under our investigation is the same as for the mechanical activation specimens (herein), as for these under natural conditions of hydration (in other studies).